\documentclass[%
 reprint,
 amsmath,amssymb,
 aps,
]{revtex4-2}
\usepackage{physics}
\usepackage{graphicx}
\usepackage{dcolumn}
\usepackage{bm}
\usepackage{float}

\begin{document}

\preprint{APS/123-QED}

\title{Observing Enhanced Entanglement in an Ising Quantum Register Beyond Weak Measurement.}%
\author{K.G.S.H. Gunawardana}
\email{harshakgs@gmail.com, korala.gunawardana@tuni.fi}
 \affiliation{Computational Physics Laboratory, Physics Unit, Faculty of Engineering and Natural Sciences, Tampere University,P.O.Box 692, FI-33014 Tampere, Finland.}





\date{\today}

\begin{abstract}
Measurements often yield a disentangled quantum state which is the most probable measurement outcome while the measurement outcomes with enhanced entanglement  could be exponentially less probable or absent.  Higher probability in measurement outcomes with enhanced entanglement is virtue in experimental observation of enhance entanglement in post-selected or Born-rule measurements with less resources in noisy quantum devices. 
We develop a periodic measurement sampling(PMS) protocol in the framework of uniform matrix  product states.  Using PMS, we identify a peak enhancement of entanglement in a uniform post-measurement state whose probabilistic outcome can be maximized by optimizing a measurement super-cell.
A measurement super-cell is the smallest repeating building block of qubits defined by the measurement density, external field strength and  exchange interactions between quantum register and ancilla qubits. 
The above prediction based on a classical description of the many-body quantum state is demonstrated in a quantum computing experiment in a loop of quantum register qubits with Ising interactions in the presence of transverse magnetic field. Further, we show the scalability of predictions of the periodic measurement sampling protocol in the thermodynamic limit to finite size systems. Especially, we present a rescaling ansatz  which would enable predicting the possibility of  observing enhanced entanglement in finite size systems.

\end{abstract}

\maketitle


\section{\label{sec:intro}Introduction}
Exploiting entanglement is the key to process quantum information  in modern noisy intermediate-scale quantum(NISQ) devices \cite{horodecki2009,eisert2010,chitambar2019}.  
The evolution under unitary gates are generic in developing entanglement while non-unitary processes such as measurements would destroy the coherent dynamics stochastically \cite{zurek2003RevModPhys,Li_PRB2019,skinner2019prx,fuji2020prb,bao2020prb,takaaki2022prl,wuPRL2026}.  
Beyond the general expectation, measurement induced entanglement(MIE) in adaptive quantum circuits would enable bypassing the slow unitary evolution and preparing highly entangled states in minimal circuit depths 
\cite{PhysRevApplied.23.054066, b7sb-nhjq, hoke_measurement-induced_2023,4b99-xmqn,PRXQuantum.3.040337,FossFeig2023uew,Buhrman2024statepreparation, iqbal_topological_2024,zi2025constantdepthquantumcircuitsarbitrary,PRXQuantum.6.010356,PhysRevX.15.021059}. 
The implementation of shallow quantum circuits is crucial in mitigating errors and reducing decoherence in NISQ-era quantum computing. 
The measurement-based quantum computation\cite{oneway2001, mbqc2003,PhysRevLett.86.5188,briegel_measurement-based_2009}, teleportation \cite{chou_deterministic_2018,gottesman_demonstrating_1999} and syndrome extraction in error correction \cite{PhysRevA.52.R2493,PhysRevA.55.900, chiaverini_realization_2004, ofek_extending_2016, Rosen_science2018, livingston_experimental_2022} are also protocols based on measurement that would potentially revolutionize quantum technologies.  
The recent studies of MIE in many-body ground state predicts enhancement of entanglement depending on the entanglement structure of the initial state \cite{Lin2023probingsign,chengPRB2024,Paviglianiti2024enhanced}.
On quantum critical ground states, measurements would result in highly nontrivial effects on correlations and entanglement depending on the measurement basis and the outcome \cite{garrattPRX2023,weinsteinPRB2023,YangPRB2023,LeePRX2023,murcianoPRX2023,khannaPRL2026,22sj-vrrn}.

 Unlike the classical observables, the highly stochastic nature of quantum measurement outcomes place a significant resource overhead and uncertainty in observing the desired outcome in measurement based computing protocols \cite{gflow2007,error2024,fbqc2025,SPflow2026} . 
Thus, higher probabilistic yield in a desired measurement outcomes is beneficial reducing computational overhead and  error mitigation in NISQ devices. However, such a deterministic control in quantum theory always comes in additional cost which would immediately follows from the Heisenberg uncertainty principle. For instance, increasing the probabilistic yield of a single outcome, due to repeated measurement or decoherence, of a multi dimensional superposition of state would result in reducing the entanglement \cite{oneway2001, mbqc2003}.

In this work, we study MIE following a single measurement conditioned on the ancilla measurement outcomes on an initially prepared ground state. 
The key question that we address is whether the probability of measurement outcomes giving enhance entanglement can be increased without compensating the entanglement gain. We find the answer is affirmative.
First, we study the problem in thermodynamic limit using the proposed periodic measurement sampling protocol. 
We find that one of the uniform measurement outcomes gives a peak enhancement of entanglement entropy following a single measurement on a low entangled state.  
The probabilistic yield of the desired uniform measurement outcome can be significantly increased without compensating the entanglement gain by varying the measurement density and the exchange interactions between the ancilla and the quantum register qubits. 
Second, we demonstrate above predictions in a quantum computing experiment in a loop of qubits supporting the periodic measurement sampling protocol.

\begin{figure*}[ht]
\centerline{\includegraphics[scale=0.4]{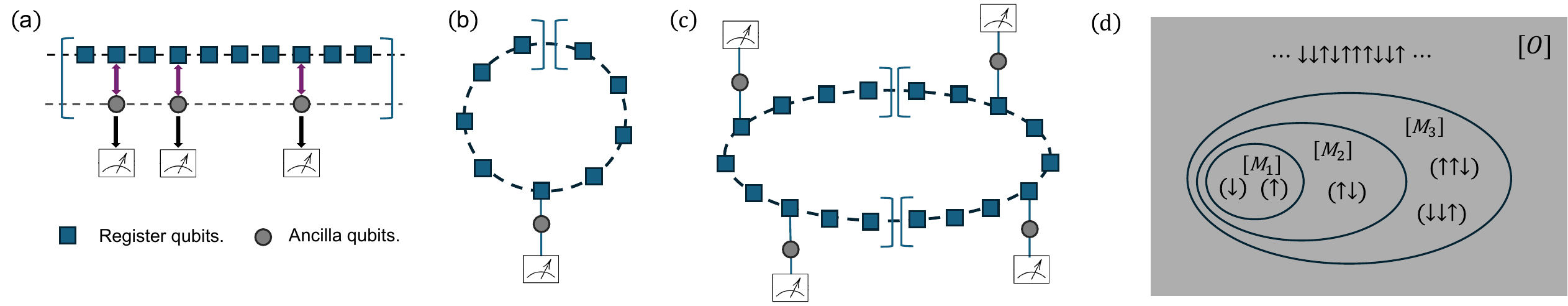}}
\caption{(a) Schematic of the measurement super-cell consisting $N_Q$ number of quantum register qubits (squares) and $N_a$ number of ancilla qubits. (b) A single measurement super-cell arranged in to a loop. (c) Periodic extension of two measurement super-cells arranged in to a loop.  (d) Periodic sampling of the measurement outcomes. $\left[M_1 \right]$,$\left[M_2 \right]$, and $\left[M_3 \right]$ are subspaces of measurement outcomes having $1,2,$ and $3$ ancilla qubits in the measurement super-cell respectively. }
\label{fig.1}
\end{figure*}

The periodic measurement sampling(PMS) protocol measures a fraction of register qubits periodically. 
The PMS protocol is implemented by introducing a measurement super-cell with $N_Q$ number of quantum register qubits and $N_a (< N_Q)$  number of ancilla qubits as depicted in Fig.\ref{fig.1} (a).  The post-measurement state of a thermodynamic limit system can be studied in a linear system having infinite number of measurement super-cells. A finite size system can also be studied by arranging single or few measurement super-cells  into a loop as shown in Fig.\ref{fig.1} (b) and (c). 
In PMS, the periodic measurement outcomes  are sampled in to subspaces as shown in Fig.\ref{fig.1} (d).  A measurement subspace consists of only the distinct measurement outcomes in a super-cell repeated through the number of super-cells in the system. 
We map the probabilities of periodic measurement outcomes in  finite size systems containing different number of super-cells into a subspace using an ansatz for post-selection.  We find that the subspace probabilities of systems having different number of super-cells would fall approximately on a single curve as that of the thermodynamic limit system. 
The above scaling would enable predicting the probabilities of  measurement outcomes in a finite size system using only the theoretically calculated subspace probabilities of an infinitely long system. 
Thus, the subspace probability could be utilized as a good parameter to optimize the  measurement super-cell properties to have higher probabilistic yield in the desired outcome.  
We discuss three major routes in optimizing the super-cell in terms of measurement density, measurement strength and initial structure of entanglement referring to Fig.\ref{fig.3}.

In proceeding, a detail description of the proposed PMS protocol is presented in section \ref{sec:PMP}. The periodic measurement outcomes in the thermodynamic limit is analyzed using tensor network calculation by applying measurement operators on the uniform matrix product state representation of an initial ground state.  The quantum computing experiment is carried out by preparing an approximate ground state using adiabatic state preparation as depicted in the  quantum circuit  in Fig. (\ref{fig.2}) for finite size loops of register qubits.  The results from both approaches are discussed  in section \ref{sec.results}. A concluding discussion is presented at the end. 

\section{\label{sec:PMP} Periodic Measurement Sampling Protocol}
 We consider measurement outcomes in a thermodynamic limit system prepared by linearly extending infinite number of measurement super-cells shown in Fig.\ref{fig.1} (a).  
There are infinite number of measurement outcomes in an infinitely long system which belongs to the infinite set $[O]$ represented by the shaded region in Fig.\ref{fig.1} (d). 
The proposed periodic measurement sampling(PMS) protocol sample the measurement outcomes in to subspaces and provides a convenient basis for numerical analysis. 
 In Fig.\ref{fig.1} (d), the closed regions, $[M_1]$, $[M_2]$, and $[M_3]$  depict resultant subspaces of measurement outcomes of a periodic system with $N_a=1$, $2$, and $3$.  
The subspace $[M_1]$ is generated when there is a single ancilla qubit in the super-cell and has only two uniform measurement outcomes denoted by $\left( \downarrow \right)$ and $\left( \uparrow \right)$.
An outcome denoted by $\left( \downarrow \right)$ represent repetitive extensions through all the super-cells in the system. For instance, when there is an infinite number of super-cells $\left( \downarrow \right) \equiv \left( \downarrow \right)^{\infty}$. The outcomes in the subspace $[M_2]$ can be denoted by $\left( \downarrow \downarrow\right)$, $\left( \downarrow \uparrow\right)$, $\left( \uparrow \downarrow\right)$ and $\left( \uparrow \uparrow\right)$ and carry the same meaning as described above. A loop of qubits as shown in Fig.\ref{fig.1} (b) and (c) can also be studied under the PMS. For a loop consists of finite number of measurement super-cells($N_{sc}$), the set $[O]$ has finite number of elements. We would follow post-selection to extract only the distinct outcomes  in a given subspace. For instance, the subspace $[M_2]$ represent the outcomes $\left( \downarrow \downarrow\right)^{N_{sc}}$, $\left( \downarrow \uparrow\right)^{N_{sc}}$, $\left( \uparrow \downarrow\right)^{N_{sc}}$ and $\left( \uparrow \uparrow\right)^{N_{sc}}$ in a finite size system. 

The total probability of the working subspace is normalized. 
The probability of the outcomes in the subspace $[M_1]$ can be written as $P_{\left( \downarrow \right)}^{[M_1]}$ and $P_{\left( \uparrow \right)}^{[M_1]}$ satisfying $P_{\left( \downarrow \right)}^{[M_1]} + P_{\left( \uparrow \right)}^{[M_1]}=1$. We could study the same system in a larger subspace by adding more ancilla spins to the super-cell. When $N_a=2$, the probabilities of the outcomes in the subspace $[M_2]$ are, $P_{\left( \downarrow \downarrow\right)}^{[M_2]}$, $P_{\left( \downarrow \uparrow\right)}^{[M_2]}$,$P_{\left( \uparrow \downarrow\right)}^{[M_2]}$ and $P_{\left( \uparrow \uparrow\right)}^{[M_2]}$ satisfying $\sum_{(m)} P_{(m)}^{[M_2]}=1$.  Thus, the probabilities of the uniform outcomes follow $ P_{\left( \downarrow \right)}^{[M_1]}> P_{\left( \downarrow \downarrow\right)}^{[M_2]}>P_{\left( \downarrow \downarrow \downarrow\right)}^{[M_3]}$.
In infinitely long systems, the subspace probabilities can be naturally extracted from the tensor network calculation following uniform Matrix Product State (uMPS) representation as given by the eq. (\ref{eq:ps_prob}).  These probabilities of the measurement outcomes always satisfy the normalization, $\sum_{(m)} P_{(m)}^{[M]}=1$ in any subspace as a direct consequence of the POVM property of the measurement operators. 

A quantum computing experiment is carried out for a loop of measurement super-cells by preparing an approximate ground state using Adiabatic State Preparation (See Fig.\ref{fig.2}).
 We find a good agreement of  entanglement restructuring in quantum register loop consisting few measurement super-cells with that of the thermodynamic limit system (See Fig. \ref{fig.5} (a) and (b)).  We follow a post-selection procedure described towards the end of section \ref{sec.results} to map the experimental measurement outcomes to the relevant subspace.  Interestingly, when the probabilities of periodic measurement outcomes $P_{(m)}$ map to a single super-cell, the post-selected subspace probabilities $P_{(m)}^{[M]}$ of systems having different number of super-cells fall approximately on to a single curve (See Fig. \ref{fig.5} (e) and (f)). Thus, as one of the main findings of this work, we establish the anzatz $P_{(m)}\approx\left(P_{(m)}^{[M]}\right)^{N_{sc}}$ to predict the probabilities of measurement outcomes in finite size systems when the subspace probabilities $P_{(m)}^{[M]}$ are known.

\subsubsection*{The System Hamiltonian and Ancilla Coupling.}
The interactions between quantum register qubits are  modeled by the Ising model in the presence of transverse magnetic field whose Hamiltonian is given as,  
\begin{equation}
    H= \sum_i -\sigma_i^z \sigma_{i+1}^z - g  \sigma_i^x, 
\label{eq:TFIM}
\end{equation}
where $\sigma^q$, $q\in \{x,y,z\}$ is the Pauli spin operators. This model shows the critical behavior when $g=1.0$. A fraction of quantum register qubits are measured periodically by coupling to ancilla qubits. There are no explicit interactions between ancilla qubits and they are initialized in a product state, $\ket{\left(\downarrow_1 \cdots \downarrow_{N_a}\right)}$. The state in side the parenthesis refers to ancilla qubits within a measurement super-cell. The states $\ket{\uparrow}$  and $\ket{\downarrow}$  are the eigenstates of the Pauli operator $\sigma^z$ and considered as the measurement basis. The quantum register and ancilla qubits are coupled via unitary interactions $U(u, \alpha)$ described by,
\begin{equation}
 U(u,\alpha)=e^{-i \left(u \sigma_{z}-\alpha \mathbb{I} \right)\otimes \widetilde{\sigma_{x}}}.
\label{eq:Uzx}
\end{equation}
The parameter $\alpha$ should be non-zero to alter the entanglement structure in the data qubits when ancilla qubits are initialized in a product state \cite{22sj-vrrn}. A weak measurement limit can be defined when $\alpha<<1$ and the weakest limit of measurement is reached near $u=1/4$ where the measurement induced back-action is minimum.  In this work, the measurement induced effect is studied beyond the above weak measurement limit in a broader range of $u$ for $\alpha\ne0$. Note the $\alpha$, and $u$ are measured in units of $\pi$. 

\subsection{Tensor network calculation.}
A periodic extension of the measurement super-cells to the thermodynamic limit is subjected to the tensor network calculation. A state of the quantum register spins can be well described by a uniform Matrix Product State . In this limit the ground state of the quantum register is numerically calculated using the Variational Uniform Matrix Product State(VUMPS) algorithm \cite{PhysRevB.97.045145,PhysRevB.97.235155,SciPostPhysLectNotes.7}.   In uniform MPS representation the ground state can be written in the form,  
\begin{equation}
    \ket{\psi_{gs}}=\sum_{s} \cdots A^{s^{n-1}}A^{s^n} A^{s^{n+1}}\cdots \ket{s}
\end{equation}
The $A^{s^n}$ is a tensor representing quantum register qubits of a super-cell and the superscripts $S^n$ represent the spin profile of $N_Q$ register qubits in a super-cell. The $A^{s^n}$ can be decomposed into a product of tensors located at each register qubit site in a super-cell as   $A^{s^n}\equiv \left(A_1^{s_1^n}A_2^{s_2^n} \cdots A_{N_d}^{s_{N_d}^n} \right)$. 
The ancilla qubits in super-cells also form an infinite long system and are initially in a product state, $\ket{\psi_a}$. 
Thus, the initial state of the infinitely long chain of  measurement super-cells can be written as,
\begin{equation}
\ket{\psi_i}=\ket{\psi_{gs}} \ket{\psi_a}. 
\end{equation}
The quantum state of the system after the unitary coupling can be expressed as , 
\begin{equation}
\ket{\psi_U}=\prod_j U_j \ket{\psi_{gs}} \ket{\psi_a},
\end{equation}
where the unitary operator $U_j$ couples the $j^{th}$ ancilla spin of the super-cell to $k^{th}$ spin of the Ising register as depicted in the Fig. \ref{fig.1} a.  A fraction of register spins are measured subjected to $N_a < N_d$.  A predetermined  set of $k$'s in a super-cell will be repeated through out the infinitely long system. 

Next the ancilla spins are measured projectively with the projector $\ket{\psi_{a'}}\bra{\psi_{a'}}$.  The state of the critical system after the projective measurement can be expressed as, 
\begin{equation}
\ket{\psi_{m}}=\frac{1}{\sqrt{p_m}}\bra{\psi_{a'}}\prod_j U_j \ket{\psi_{gs}} \ket{\psi_a},
\end{equation}
where $\ket{\psi_{m}}$ is the post measurement state of the critical system and $p_m=\bra{\psi_m} \ket{\psi_m}$ is the probability of the post-measurement state $\ket{\psi_m}$.  
Assuming the ancilla is initially in a product state, the post measurement state can be written as, 
\begin{equation}
\ket{\psi_{m}}=\frac{1}{\sqrt{p_m}}\left(\prod_j \bra{a'_j} U_j \ket{a_j}\right) \ket{\psi_{gs}},
\end{equation}
where $\ket{a_j}$ and $\ket{a'_j}$ are initial state and the measurement outcome of the ancilla spin at site $j$ respectively. 
Now the measurement operators (Kraus operators) due to the $j$th ancilla spin can be defined as, 
\begin{equation}
M_j^{a'a}=\bra{a'_j} U_j \ket{a_j}.
\label{eq:M_op}
\end{equation} 
For a given initial state of the $\ket{a}$ the $M_j^{a'a}$ satisfy the condition for a positive operator-valued measure(POVM),
\begin{equation}
 \sum_{\{a'\}}\left(M^{a'a}_j\right)^{\dagger} M^{a'a}_j=1, 
 \label{eq:povm}
\end{equation} 
which ensure the normalization of the total Born-rule probability. 

Using the equation \ref{eq:M_op}, we can derive the measurement operators, 
\begin{align}
M^{\downarrow \downarrow}(u,\alpha)=\begin{pmatrix}
\cos (u+\alpha) & 0 \\
0 & \cos (u-\alpha)
\end{pmatrix}
\end{align}
\begin{align}
M^{\uparrow \downarrow}(u,\alpha)=\begin{pmatrix}
 \sin (u+\alpha) & 0 \\
0 & -\sin (u-\alpha)
\end{pmatrix}
\end{align}
Consequently, for a super-cell with $N_a=2$, the post measurement states correspond to the  measurement outcome $m\equiv\left(\downarrow \uparrow \right)$ can be written as,  
\begin{equation}
\ket{\psi_{\left(\downarrow \uparrow \right)}}= \prod_k \left(M_1^{\downarrow \downarrow}M_2^{\uparrow \downarrow}\right)_k \ket{\psi_{gs}}, 
\label{eq:PMstate1}
\end{equation}
where $k$ refers to the super-cell index.  In the uniform-MPS, the quantum state of the infinitely long system can be uniquely express by the product of $N_Q$ number of 3-dimensional tensors in the super-cell. Thus the post measurement state within the super-cell(sc) can be expressed as, 
\begin{equation}
\ket{\psi^{sc}_{\left(\downarrow \uparrow \right)}}=  \left(M_1^{\downarrow \downarrow}M_2^{\uparrow \downarrow}\right) \ket{\psi^{sc}_{gs}}, 
\label{eq:state_sc}
\end{equation}
where the $\ket{\psi^{sc}_{gs}}$ is the ground state defined by the super-cell tensors. The probabilities of measurement outcomes in a given subspace  ( As shown in Fig. \ref{fig.1}d) can be numerically calculated  by, 
\begin{equation}
    P^{[M]}_{(m)}=\bra{\psi^{sc}_{(m)}} \ket{\psi^{sc}_{(m)}}. 
    \label{eq:ps_prob}
\end{equation}
The $(m)$ is a $N_a$ dimensional vector representing a measurement outcomes in the subspace.  In a given subspace the post selected probabilities, satisfies the normalization condition, $\sum_{(m)} P^{[M]}_{(m)}=1$, which is guaranteed by the POVM property of the measurement operators. In weak measurement limit where $\alpha << 1$, the approximate post selected probabilities can be analytically expressed by the terms of the binomial expansion $\left(\cos^2(u)+\sin^2(u)\right)^{N_a}$ ($=1$). The powers of the terms $\cos^2(u)$ and $\sin^2(u)$ refer to number of $\downarrow$ and $\uparrow$ states in the measurement outcomes in ancilla qubits.  The  expansion coefficients account for the degeneracy. See the reference \cite{22sj-vrrn} for more details on the weak measurement limit.  Beyond the weak measurement limit the  probabilities of each outcome deviate drastically from the that of weak measurement limit but satisfy the normalization condition.

\subsection{Quantum computing experiment.\label{sec:Quantum Experiment}}
\begin{figure}[ht]
\centerline{\includegraphics[scale=0.4]{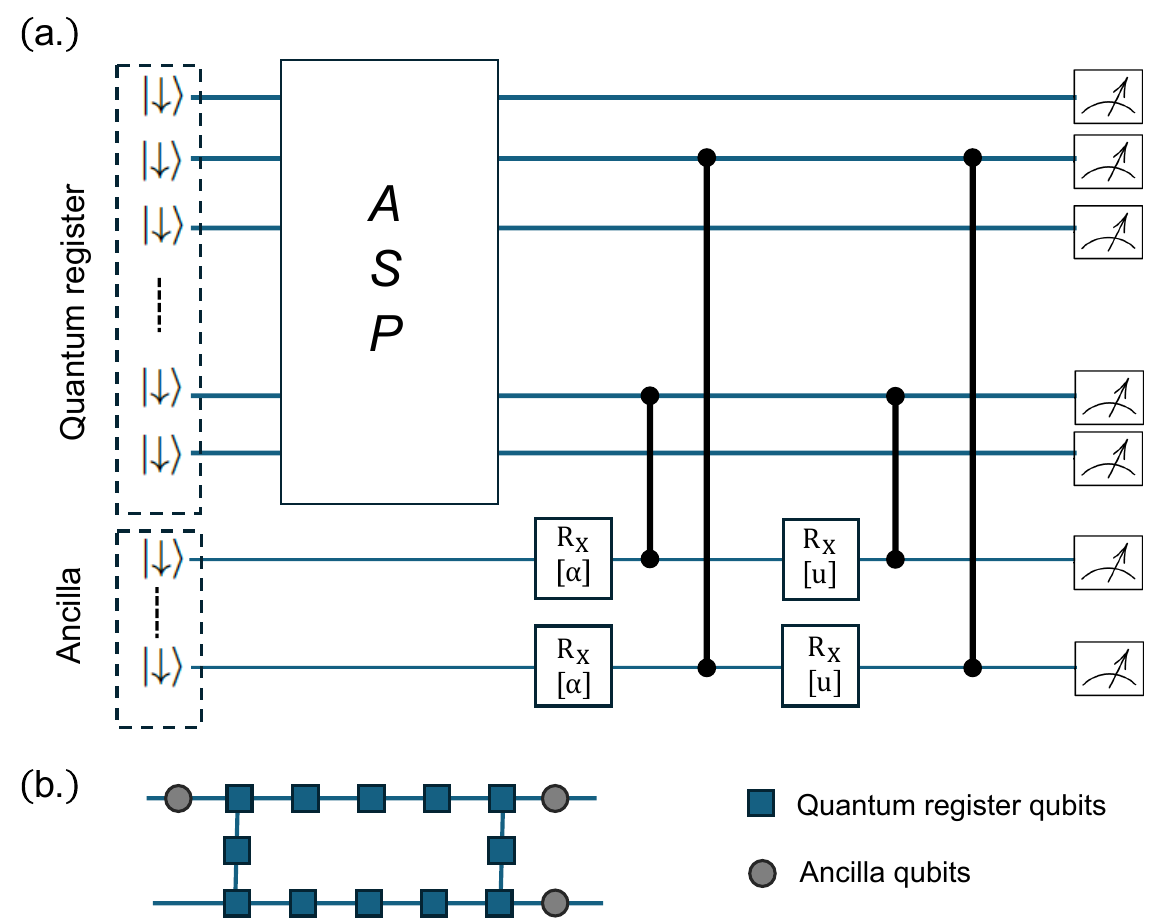}}
\caption{(a.) Quantum circuit. The adiabatic state preparation(ASP) block consists of first-order trotter steps following the time dependent evolution in eq.(\ref{eq:time_Evol}). The single qubit gate $R_X[\theta]$ rotate the spin vector on $x-$axis by an angle $\theta$ and the dotted vertical connectors represent $2-$qubit CZ gates. (b.) Quantum processor layout. A single measurement super-cell is defined with periodic boundary condition in the ibm-kingston quantum processor. The super-cell contains $12$ quantum register qubits and $3$ ancilla qubits.  }
\label{fig.2}
\end{figure} 

We implement Adiabatic State Preparation(ASP) algorithm to prepare an approximate ground state of a loop of quantum register qubits consisting finite number of super-cells.  In ASP the system is initialized in the ground state of an easily preparable Hamiltonian($H_i$) and slowly evolve to the target Hamiltonian($H_f$). The quantum adiabatic theorem guarantees that the system follows through instantaneous ground states given a sufficiently slow evolution to the target Hamiltonian \cite{RevModPhys.90.015002, PhysRevA.105.032403, Fuller17012026}.  We can define a time dependent Hamiltonian,
\begin{equation}
     H_{ad}\left(t\right)=\left(1-S\left(\frac{t}{T}\right)\right)H_{ini} +S\left(\frac{t}{T}\right) H_f,
    \label{eq:Adiab_H}
\end{equation}
where the target Hamiltonian is the TFIM Hamiltonian given in eq.(\ref{eq:TFIM}), $T$ is the total evolution time and $S(t/T)$ is the smooth scheduling function satisfying $S(0)=0$ and $S(1)=1$.  The ASP is initialized with the Hamiltonian, 
\begin{equation}
    H_{ini}=-\sum_i \sigma^z_i \sigma^z_{i+1},
\end{equation}
whose the ground state is $\ket{\psi_{gs}(0)}=\ket{\downarrow \downarrow \cdots \downarrow}$. The desired ground state of the Hamiltonian ($\ref{eq:TFIM}$) is obtained by the time evolution, 
\begin{equation}
    \psi_{gs}(1)=\mathcal{T}\exp\left(-i\int_0^TH(t)dt\right) \ket{\psi_{gs}(0)}, 
    \label{eq:time_Evol}
\end{equation}
where $\mathcal{T}$ denotes the time-ordering operator. We adapt the scheduling function $S(t/T)=\sin(\pi t/2T)$ through out this work and implement a first order trotter algorithm to decompose the unitary operator in Eq. (\ref{eq:time_Evol}) at a given instance into two-qubit gates. 

The prepared approximate ground state of the quantum Ising resister is coupled to set of ancilla qubits as shown in the Fig.\ref{fig.2} (a.).   The unitary interactions $U(u,\alpha)$ given in the equation (\ref{eq:Uzx}) can be equivalently implemented in the quantum circuit using the rotational (${R_X[\theta]}$) operator  and the control-z gates($CZ$) such that, 
\begin{equation}
 e^{-i \left(u \sigma_{z}-\alpha \mathbb{I} \right)\otimes \widetilde{\sigma_{x}}} \equiv e^{i\alpha \sigma_x}. CZ \left(\mathbb{I}\otimes e^{-i u \sigma_x}\right) CZ.
\label{eq:RCZ-couple}
\end{equation}
The quantum register qubits outcomes are analyzed by post-selecting on ancilla qubits after sufficient number of  measurement outcomes are collected by repeating the quantum experiment.  

\section{Results \label{sec.results}}
\begin{center}
\begin{figure*}[ht]
\centerline{\includegraphics[scale=0.35]{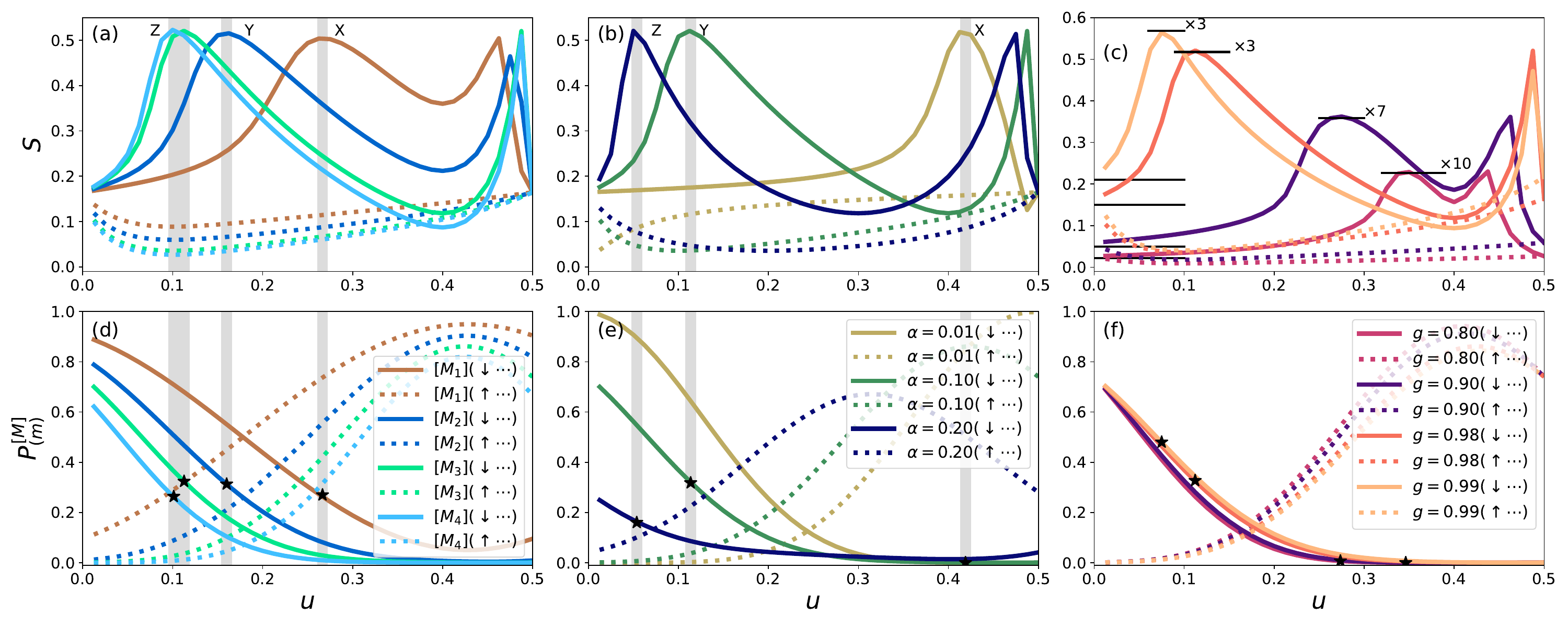}}
\caption{ Entanglement entropy(S) and subspace probability, $P_{(m)}^{[M]}$ of the measurement outcome $(m)$. The solid lines refer to the measurement outcomes $\left(\downarrow \cdots \right)$ in the corresponding subspace while the dotted lines refer to the measurement outcome $\left(\uparrow \cdots \right)$ . (a) Entanglement entropy $S$ against the unitary coupling parameter $u$ at $\alpha=0.1$ in different measurement subspaces $\left[M_1 \right]$,$\left[M_2 \right]$, $\left[M_3 \right]$ and  $\left[M_4 \right]$. The measurement density in each subspace is $10\%$, $20\%$,$30\%$,and $33\%$ respectively. (b) $S$ vs. $u$ for different measurement strength ($\alpha$) calculated at $g=0.98$ in subspace $\left[M_3 \right]$. (c) $S$ vs. $u$ at different field strength of transverse magnetic field $g$ in the subspace $\left[M_3 \right]$ at $\alpha=0.1$. The approximate gain in $S$ is marked by the peak of $S$. (d), (e), (f) The variation of $P_{(m)}^{[M]}$ for the cases shown in (a), (b) and (c) respectively. The $\star$ mark shows the $P_{(m)}^{[M]}$ at the peak of $S$.  }
\label{fig.3}
\end{figure*}
\end{center}
\subsubsection{Tensor network calculation for infinite systems.}
The ground state of the Hamiltonian (\ref{eq:TFIM}) is calculated in the thermodynamic limit using the VUMPS algorithm keeping $N_Q$ register qubits in the super-cell.  The post measurement state in the super-cell is calculated by applying the measurement operators as  given in eq. \ref{eq:PMstate1}. We observe that the bond dimension $\chi=32$ is sufficient to accurately obtain the ground state near the critical point $g=1.0$ in a manageable computation cost. The von-Neumann entanglement entropy ($S$) is calculated using singular value decomposition (SVD) of approximated finite-size MPS of the post measurement state. 

Fig. \ref{fig.3} (a) shows the variation of S as a function of $u$  in subspaces $\left[M_1 \right]$, $\left[M_2 \right]$, $\left[M_3 \right]$, and $\left[M_4 \right]$, when $g=0.98$ and the strength of measurement $\alpha=0.1$. The subspaces $\left[M_1 \right]$, $\left[M_2 \right]$, and $\left[M_3 \right]$ have $10$ register qubits in the super-cell and $\left[M_4 \right]$ has $12$ register qubits. Thus the measurement density in each subspace is $10\%$, $20\%$, $30\%$, and  $33\%$ respectively. The Fig.\ref{fig.3} (d) shows the variation of the subspace probability, $P_{(m)}^{[M]}$ as a function of $u$ calculated from the uniform MPS. 
The uniform measurement outcomes in a given subspace is denoted by $(\downarrow \cdots )$ and  $(\uparrow \cdots )$ for the convenience. 
The solid lines  represent the outcome $(\downarrow \cdots )$  and the dotted lines represent the outcome $(\uparrow \cdots )$ given that the ancilla qubits are initiated in the $(\downarrow)$ state. The $S$  of the outcome $(\downarrow \cdots )$ increases significantly and shows unique non-monotonic behavior in the domain $0<u<0.5$ (note the angles are measured in unit of $\pi$). On the other hand $S$ of the outcome $(\uparrow \cdots )$ decreases similar to the general expectation of projective measurements. Thus, the main discussion in this work is focused on the measurement outcome   $(\downarrow \cdots )$ showing the anomalous restructuring of entanglement. 

The $S$ of the post measurement state $(\downarrow \cdots )$ shows two distinct peaks. The peak entanglement entropy is approximately 3 times that of the initial ground state and is nearly a constant for different sub-spaces.  The location of the first broader peak in each curve is marked in shaded regions  $X$, $Y$ and $Z$ in Fig. \ref{fig.3} (a). In an experiment  choosing the unitary coupling parameter $u$ on peak position in $X$, $Y$ and $Z$ regions for relevant measurement density or sub-space  would give the maximum enhancement of $S$ of the post measurement state.    
The subspace probability at the peak of the entanglement entropy is denoted by $\star$ in Fig. \ref{fig.3}(d). The highest probability of the outcome  $(\downarrow \cdots )$ occurs when $u\rightarrow 0$, at which no change of S is observed. At this limit the ancilla qubits disentangle from the quantum register. 

In the subspace $[M_1]$, the probability of the $(\uparrow \cdots )$ dominates that of $(\downarrow \cdots )$ at the peak of the entanglement entropy( see the shaded region x). Note, there are only two measurement outcomes in the subspace $[M_1]$ such that $P_{(\downarrow\cdots)}^{[M_1]}+P_{(\uparrow\cdots)}^{[M_1]}=1$.  The measurement density can be increased by adding more ancilla qubits to the super-cell so that the measurement outcomes lies in a larger subspace increasing the number of distinct measurement outcomes. Thus, an overall reduction of the probability of an individual measurement outcome  can be expected. However, we observe a slight increase of the probability of the outcome $(\downarrow \cdots )$  when moving to the subspaces $[M_2]$ and $[M_3]$. See the $\star$ mark in the shaded region $Y$ and $Z$. The probability of the outcome $(\uparrow \cdots )$ reduces significantly when $u\rightarrow 0$.  Moreover, further increasing the measurement density the probability of the outcome $(\downarrow \cdots )$ decreases. See the first $\star$ mark from the left. Thus, the PMS protocol predicts that there could be an optimal measurement density to observe a peak entanglement entropy with relatively higher probability. For the used case at $\alpha=0.1$, the measurement density $30\%$ could be near optimum at  $u\approx 0.1$. 

The Fig. \ref{fig.3}(b) shows the dependence of $S$ on the measurement strength $\alpha$ in the subspace $\left[M_3\right]$ for $g=0.98$. In weak measurement $\alpha=0.01$, the peak of $S$ labeled as $X$ have negligibly small subspace probability as denoted by $\star$ in Fig. \ref{fig.3} (e).   Thus, experimental observation of enhanced entanglement could be exponentially less probable even in a system with a single super-cell.  When $\alpha$ increases beyond the weak limit, the position of the peak entanglement moves towards lower values of $u$ having higher probability.  The subspace probability of observing the peak of $S$ is about $30\%$ at $\alpha=0.1$.  However, further increment of $\alpha$ significantly lower the subspace probability. Note the $\star$ mark in the shaded region $Z$ in Fig.\ref{fig.3}(e) for $\alpha=0.2$ .   

The Fig. \ref{fig.3} (c), shows the variation of $S$   of uniform measurement outcomes at different field strengths of the transverse magnetic field $g$ in the subspace $\left[M_3\right]$ at $\alpha=0.1$. Our study is focused on the magnetic phase of the Hamiltonian (\ref{eq:TFIM}) away from the critical point, $g<1$.   
Note the $S$ of the initial ground state  at $g=0.98$ before the measurement is $S_{init}\approx0.15$ (horizontal line segment on the left of Fig. \ref{fig.3} (c)). We observe about $3-$fold enhancement of $S$ following a single measurement. 
The observed peak enhancement of $S$ at $g=0.90$ and $g=0.80$ is about $7-$fold and $10-$fold respectively.  As shown in Fig. \ref{fig.3} (f), the variation of $P_{(m)}^{[M]}$ has weak dependence on $g$ while the $S$ shows a significant variation with $g$. Note the $P_{(m)}^{[M]}$ at the peak of $S$ decreases drastically  when $g$ decreases. 
Thus, observing the $7-$fold and $10-$ fold enhancement of $S$ is comparatively less probable. See the first two $\star$ mark from the right in Fig. \ref{fig.3} (f).  
The $P_{(m)}^{[M]}\approx 0.35$ when $g=0.98$ at the peak of $S$. See the 3rd $\star$ mark from the right in Fig. \ref{fig.3} (f). 
The $P_{(m)}^{[M]}$ at the peak of $S$ increases further when $g$ increases. In the tensor network calculation we observe that  $P_{(m)}^{[M]}\approx 0.5$  at $g=0.99$. Moreover, the the gain of $S$ is about $3-$fold at the peak. 
The $S$ of the initial ground state is $S_{init}\approx0.21$ at $g=0.99$. 
Thus, the field strength $g=0.99$ can be considered as relatively away from the critical point. Rapid divergence of the entanglement entropy occurs further closer to the critical point. 
We would anticipate that  $P_{(m)}^{[M]}> 0.5$ at the peak of $S$ when $g>0.99$. Thus, the measurement outcome giving enhance entanglement would be the dominating measurement outcome in the subspace near the critical point.   

\subsubsection{Quantum computing experiment for finite size systems.} 
\begin{center}
\begin{figure*}[ht]
\centerline{\includegraphics[scale=0.35]{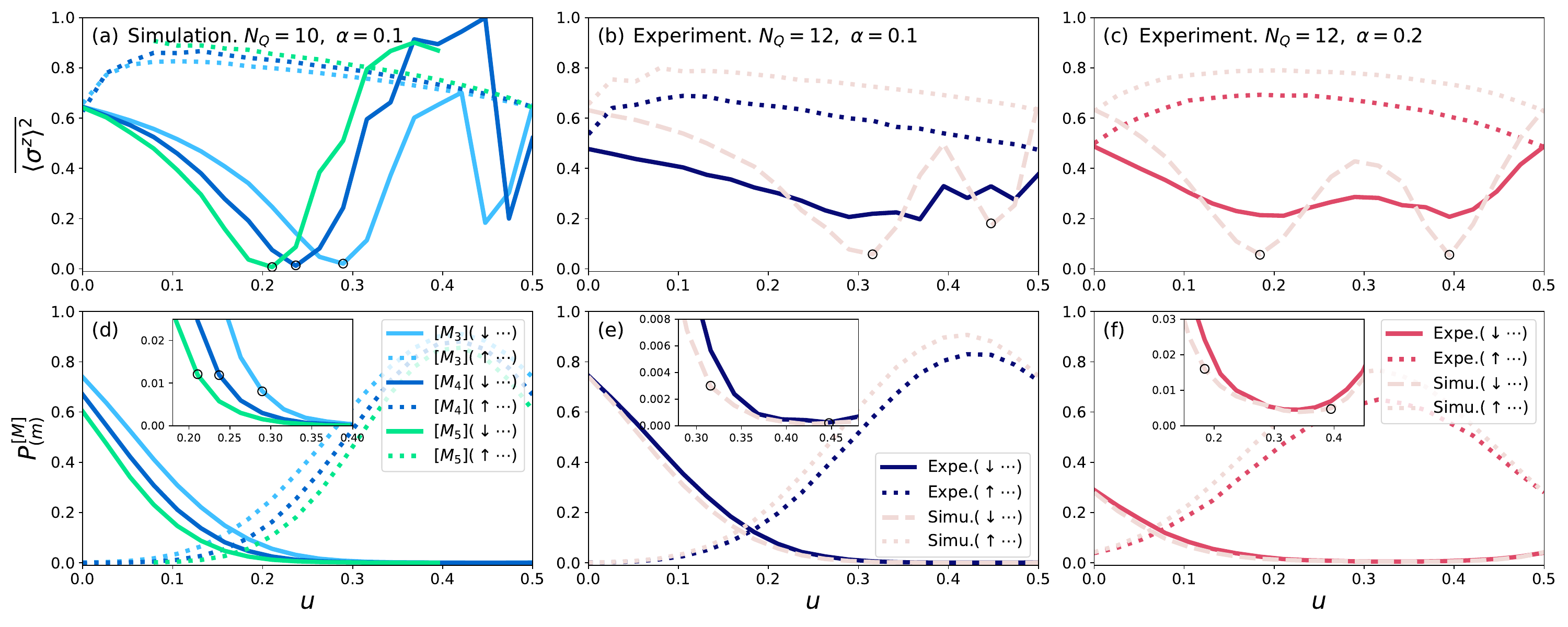}}
\caption{ The mean square magnetization, $\overline{\langle \sigma^z \rangle^2}$ and subspace probabilities, $P_{(m)}^{[M]}$ of uniform measurement outcomes of approximate ground state prepared from ASP. (a) Variation of $\overline{\langle \sigma^z \rangle^2}$ vs. $u$ from simulated experiment using Qiskit aer simulator calculated in $[M_3]$, $[M_4]$, and $[M_5]$ subspaces. The minima of $\overline{\langle \sigma^z \rangle^2}$ correspond to the maximum entanglement of the post-measurement state $(\downarrow \cdots)$.  (b) \& (c) Results from the quantum computing experiment conducted in the ibm-kingston quantum processor.  The super-cell contains $12$ quantum register qubits and $3$ ancilla qubits as shown in Fig.\ref{fig.2} (b). The results are produced using measurement outcomes within $5\times 10^4$ shots. The curves in light colors represent the simulated experiment for an identical system using $5\times 10^5$ shots. (d),(e) \& (f) $P_{(m)}^{[M]}$ as a function of $u$ correspond to the results shown in (a), (b) \& (c) respectively. The insets shows a closeup view of the minima of $\overline{\langle \sigma^z \rangle^2}$ marked by open circles. }
\label{fig.4}
\end{figure*}
\end{center} 

In the experiment, an approximate ground state is prepared in a quantum processor following the adiabatic state preparation(ASP) and a fraction of qubits is measured periodically coupling to ancilla qubits as described in the section \ref{sec:Quantum Experiment}. Fig. \ref{fig.2} (b) shows the layout of the  loop of qubits used in the ibm-kingston quantum processor. The qubits in the loop consisting 12 quantum resister qubits are coupled to 3 ancilla qubits. Thus, the system is a single measurement super-cell in the subspace $[M_3]$. 
In the ASP, the trotter timestep for the chosen depth of the circuit is determined by minimizing the energy expectation value of the Hamiltonian (\ref{eq:TFIM}). 
Once the optimal trotter step is determined the ASP is repeated and proceed to the measurement as shown in  the quantum circuit in Fig.\ref{fig.2} (a).

To quantify the entanglement restructuring, we estimate the average local magnetization of post-selected uniform outcomes in the experiment. We define mean-square local magnetization \cite{PhysRevB.110.054207} as, 
\begin{equation}
    \overline{\langle \sigma^z \rangle^2} = \frac{1}{N_Q}\sum_{i=1}^{N_Q} \langle \sigma^z_i \rangle^2.
\end{equation}
First, the above proposed experiment is simulated using Qiskit Aer simulator. The simulation is conducted for a  quantum register loop of $N_Q=10$ described by the Hamiltonian eq.($\ref{eq:TFIM}$) with $g=0.90$. The Fig.\ref{fig.4} (a) shows the calculated $\overline{\langle \sigma^z \rangle^2}$ as a function of $u$ for the uniform measurement outcomes of the simulated experiment. The measurements are done in the $[M_3]$, $[M_4]$, and $[M_5]$ subspaces. Each curve is constructed collecting measurement outcomes of the quantum circuit over $1\times10^5$ shots. The $\overline{\langle \sigma^z \rangle^2}$ profiles for  both uniform measurement outcomes show the characteristic features analogues to the $S$ which can be qualitatively described by $\overline{\langle \sigma^z \rangle^2} \propto 1/S $.  
For the $[M_3]$ subspace at $\alpha=0.1$, the location ($u$ value) of the two minima of the outcome $(\downarrow \cdots)$ of  $\overline{\langle \sigma^z \rangle^2}$ are reasonably closer to the peaks of $S$ at $g=0.90$ shown in the Fig.\ref{fig.2}(c). Thus, we could consider this fact as an indirect observation to verify that  the simulated ASP ground state is sufficiently close to the VUMPS ground state despite the finite size of the system used in ASP. The location of the first minima moves toward low values of $u$ when increasing the number of ancilla qubits. Note the data for the measurement outcome $(\downarrow \cdots)$ for $[M_4]$ and $[M_5]$. Further, the subspace probability $P_{(m)}^{[M]}$ depicted in  Fig.\ref{fig.4} (d) decreases when increasing the measurement density similar to the prediction from the VUMPS calculation shown in Fig.\ref{fig.3} (d).  
In the subspace $[M_5]$, there is not a single measurement outcome recorded within the $1\times 10^5$ shots in the regions $u>0.4$ for $(\downarrow \cdots)$ and $u<0.1$ for $(\uparrow \cdots)$. The slightly elevated $P_{(m)}^{[M]}$ near $u\approx 0.2$ collects sufficient statistics to observe the minimum of $\overline{\langle \sigma^z \rangle^2}$ in the $[M_5]$. The $P_{(m)}^{[M]}$ at the first minima of $\overline{\langle \sigma^z \rangle^2}$  increases slightly when increasing the measurement density as shown in the inset of Fig. \ref{fig.4} (d), which is consistent with the thermodynamic limit observation shown in Fig.\ref{fig.3} (d). 

In the quantum computing experiment a  shallow circuit is implemented in ASP with optimized trotter step.  The depth of the ASP circuit in the experiment is only $10$ trotter steps. It is noted that the reduction of the circuit depth is crucial in mitigating errors and qubit decoherence. 
Fig. \ref{fig.4} (b) shows the variation of $\overline{\langle \sigma^z \rangle^2}$ in the experiment using ibm-kingston QPU for $g=0.90$ and $\alpha=0.1$. A loop of qubits with $N_Q=12$ is taken as the quantum register and the measurement is done in $[M_3]$. The light color curves(dashed and dotted) represent the data from the simulated experiment for the same super-cell.  
The measurement density is lower than that of the simulated experiment shown in Fig. \ref{fig.4} (a). Thus, we notice the first minima of the outcome $(\downarrow \cdots)$ is slightly shifted to the right(marked by a open circle).  The experimental results shows a similar variation for both uniform measurement outcomes. The location of the minima are well contained in the simulated minima. The reduction of the  $\overline{\langle \sigma^z \rangle^2}$ in the limit of $u\rightarrow 0$ and $u\rightarrow0.5$ could be a result of random spin flip errors and measurement errors. The $\overline{\langle \sigma^z \rangle^2}$ at $\alpha=0.2$ shown in Fig. \ref{fig.4} (c) also shows the expected variation similar to the tensor network calculation. The minima of the outcome $(\downarrow \cdots)$ are moved left and well contained in the region of the simulated minima(marked by open circles). The $P_{(m)}^{[M]}$  of the outcome $(\downarrow \cdots)$  is also in good agreement with the simulated probabilities. See Fig. \ref{fig.4} (e) and (f). The experiment is only ran for $5\times 10^4$ shots where as the simulation is ran for $5\times 10^5$ shots.
The $P_{(m)}^{[M]}$ of the outcome $(\downarrow \cdots)$ when $\alpha=0.1$ is considerably small where the minima of the $\overline{\langle \sigma^z \rangle^2}$ is located $u>0.3$. Note the inset of the Fig. \ref{fig.4} (e). This results noisy output of $\overline{\langle \sigma^z \rangle^2}$ in the relevant region due to the small number of measurement outcomes. The simulation predicts about $\sim 150$ number of $(\downarrow\cdots)$ outcomes near the first minima in the experiment.  The probability at the first minima increases by approximately  $\times5$ when the measurement strength is increased to $\alpha=0.2$. See the inset of the  Fig. \ref{fig.4} (f).  Thus the number of desired outcomes will be increased to $\sim750$ and we can observe much smoother and clear minima in $5\times 10^4$ shots in the experiment.

\begin{center}
\begin{figure}[ht]
\centerline{\includegraphics[scale=0.35]{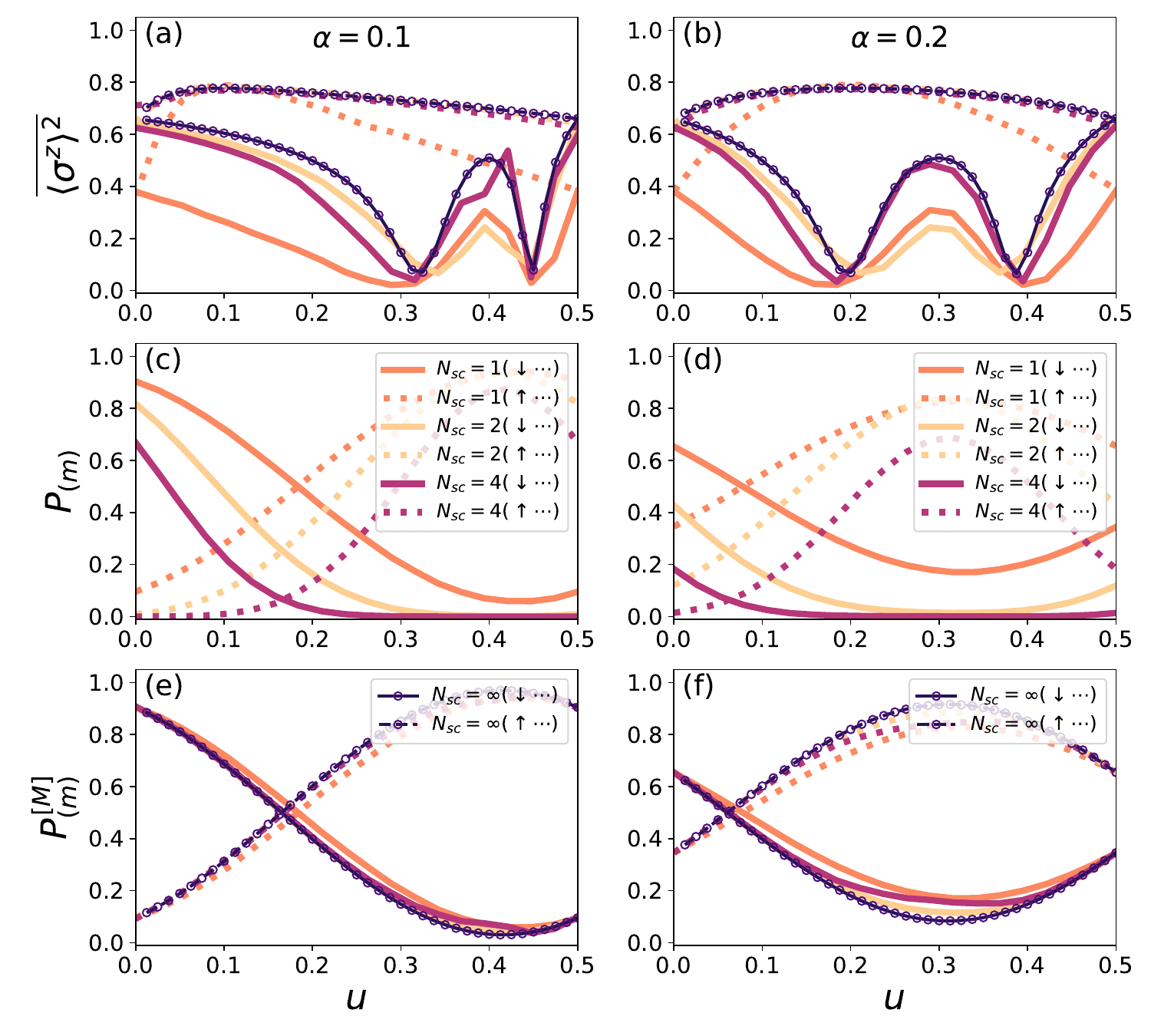}}
\caption{  Entanglement restructuring of quantum register loops consisting $1$,$2$, and $4$ super-cells. (a) \& (b) Variation of $\overline{\langle \sigma^z \rangle^2}$ calculated at $\alpha=0.1$ and $0.2$ respectively.  (c) \& (d) Probability of uniform measurement outcomes of finite size systems. (e) \& (f) Scaled probabilities of measurement outcomes of systems having different number of super-cells to the subspace $[M_1]$ following the post-section procedure described in the main text. Note: The lines with circles are calculated in tensor network  for an infinitely long system($N_{sc}=\infty$) having identical super-cell. }
\label{fig.5}
\end{figure}
\end{center}

\subsubsection{Finite size systems and scaling of probabilities.}
To analyze the finite size effect and scalability of subspace  probabilities, we study finite number of  measurement super-cells arranged in loops. We consider systems consisting $1$, $2$, and $4$ super-cell loops in $[M_1]$ subspace with $N_Q=5$.  The ground state is prepared using ASP in Qiskit Aer simulator. The Fig. \ref{fig.5} (a) and (b) shows the variation of $\overline{\langle \sigma^z \rangle^2}$  of uniform post-measurement states for $\alpha=0.1$ and $0.2$ respectively when $g=0.90$. The variation of $\overline{\langle \sigma^z \rangle^2}$ for $N_{sc}=1,2 ,$ and $4$ shares qualitatively similar features. The results for a system with infinite number of  super-cells ($N_{sc}=\infty$) in the same subspace is shown as a benchmark(black lines with circles). We assume that the results of finite  $N_{sc}$ would asymptotically close to that of $N_{sc}=\infty$  at sufficiently large  $N_{sc}$. Further, there would also be some deviations due to the approximate ground state prepared in ASP. 

The Fig.\ref{fig.5} (c) and (d) shows the probabilities of the finite size system $P_{(m)}$ of uniform measurement outcomes $(\downarrow \cdots)$ and $(\uparrow \cdots)$. The $P_{(m)}$ decreases exponentially when increasing the $N_{sc}$. Now, we scale the probabilities of different system sizes to the subspace $P_{(m)}^{[M]}$ adapting the following procedure of post-selection. 
\begin{enumerate}
    \item First post-select periodic measurement outcomes which are distinct in the measurement super-cell. For instance in $[M_3]$, there are only 8 measurement outcomes of the form $(\downarrow\downarrow \downarrow)^{N_{sc}}$, $(\downarrow\downarrow \uparrow)^{N_{sc}}$, $(\downarrow\uparrow \uparrow)^{N_{sc}}$, $\cdots$ ect. In the used case there are only two periodic outcomes in $[M_1]$, namely: $(\downarrow)^{N_{sc}}$ and  $(\uparrow)^{N_{sc}}$.
    \item Next the probabilities are mapped to a single super-cell using the ansatz $(P_{(m)})^{\frac{1}{N_{sc}}}$. 
    \item The subspace probability is calculated as, 
    \begin{equation}
        P_{(m)}^{[M]}=\frac{(P_{(m)})^{\frac{1}{N_{sc}}}}{\sum_{(m')} (P_{(m')})^{\frac{1}{N_{sc}}} }.
    \end{equation}
    The summation over $(m')$ runs over only the distinct periodic outcomes in the subspace as listed in step (1). 
\end{enumerate}
Fig. \ref{fig.5} (e) and (f) show the $P_{(m)}^{[M]}$ calculated following the above procedure. The $P_{(m)}^{[M]}$ of different system sizes falls approximately on to a single curve.  The deviation of $P_{(m)}^{[M]}$ between different system sizes increases slightly when increasing $\alpha$. 
However, $P_{(m)}^{[M]}$ of $N_{sc}=4$ and the $N_{sc}=\infty$ shows good agreement in a broader part of the curve at $\alpha=0.1$. Interestingly, the agreement at $\alpha=0.2$ is also noteworthy. Based on the numerical evidence, we would expect the curves at finite $N_{sc}$ to collapse into $N_{sc}=\infty$ at sufficiently larger $N_{sc}$. 

The above scaling  is exact in the limit $\alpha<<1$, which can be shown analytically using the post-measurement state in the limit $\alpha \rightarrow 0$. It also results $\sum_{(m')} (P_{(m')})^{\frac{1}{N_{sc}}}=1$ and hence the $P_{(m)}^{[M]}= P_{(m)})^{\frac{1}{N_{sc}}}$.
For the simulated experiment in Qiskit aer, we find that $\sum_{(m')} (P_{(m')})^{\frac{1}{N_{sc}}}\approx 1$ for all $\alpha\ne0$. This condition is tested for the results shown in Fig. \ref{fig.5} and also in subspace $[M_3]$. Thus, we can establish a rescaling ansatz for the probability of a measurement outcome in a finite size system as, $P_{(m)} \approx  \left(P_{(m)})^{[M]}\right)^{N_{sc}}$.

\section{Conclusions}
The periodic measurement sampling(PMS) protocol presented in this work provides a unified way to study measurement induced effect in different length scales.  We show that entanglement restructuring and probability of ancilla measurement outcomes can be known at different length scale systems only by studying those in measurement subspace. Further, we present a methodology to evaluate subspace properties numerically in the thermodynamic limit using uniform matrix product state representation. 
 
 In this work, we provide insight on observing the peak entanglement entropy of the post measurement state with relatively higher probability with respect to the disentangling  uniform outcomes. We demonstrate possible pathways to observe the peak entanglement with significantly higher probability. We provide experimental results  supporting the above proposal in a superconducting quantum processor. There could be an optimal measurement density and measurement strength that enable observing the peak entanglement in uniform post-measurement state with significantly higher probability. Moreover, the subspace probability at the peak entanglement would strongly depend on the amount of entanglement in the initial state. The low entanglement entropy in initial state would significantly lower the subspace probability at the peak of the entanglement in post-measurement state. By preparing the initial ground state closer
  to the critical point would increase the subspace probability significantly keeping the higher entanglement gain. Thus,we believe that these findings following PMS sheds light on creating measurement induced deep entangled state in systems having large number of qubits where classical computation can not reach.

\begin{acknowledgments}
I wish to acknowledge Teemu Ojanen for the assistance provided and the TCSC computing cluster at Tampere University. 
\end{acknowledgments}

\bibliography{PMS}

\end{document}